# *TeamTat*: a collaborative text annotation tool

Rezarta Islamaj[1,§], Dongseop Kwon[2,§], Sun Kim[1] and Zhiyong Lu[1,*]

[1] National Library of Medicine, National Institutes of Health, Bethesda, MD 20894, USA
[2] School of Software Convergence, Myongji University, Seoul 03674, South Korea

* To whom correspondence should be addressed. Tel: +1 301 594 7089; Fax: +1 301 480 2290; Email: zhiyong.lu@nih.gov

## ABSTRACT

Manually annotated data is key to developing text-mining and information-extraction algorithms. However, human annotation requires considerable time, effort and expertise. Given the rapid growth of biomedical literature, it is paramount to build tools that facilitate speed and maintain expert quality. While existing text annotation tools may provide user-friendly interfaces to domain experts, limited support is available for image display, project management, and multi-user team annotation. In response, we developed *TeamTat* (https://www.teamtat.org), a web-based annotation tool (local setup available), equipped to manage team annotation projects engagingly and efficiently. *TeamTat* is a novel tool for managing multi-user, multi-label document annotation, reflecting the entire production life cycle. Project managers can specify annotation schema for entities and relations and select annotator(s) and distribute documents anonymously to prevent bias. Document input format can be plain text, PDF or BioC, (uploaded locally or automatically retrieved from PubMed/PMC), and output format is BioC XML with inline annotations. *TeamTat* displays figures from the full text for the annotator's convenience. Multiple users can work on the same document independently in their workspaces, and the team manager can track task completion. *TeamTat* provides corpus-quality assessment via inter-annotator agreement statistics, and a user-friendly interface convenient for annotation review and inter-annotator disagreement resolution to improve corpus quality.

## INTRODUCTION

Gold-standard corpora, collections of text documents semantically annotated by domain experts, are crucial for the development and training of text-mining and information-extraction algorithms. Particularly in the life sciences, where texts are full of biomedical entities whose naming often does not follow convention and the relationships between entities may differ in subtle ways (1-8), annotation tools need to provide support for multiple domain experts to review and annotate, for automatic annotation comparisons, as well as for the tracking of annotation consistency. These capabilities will allow one to identify relevant differences in annotation patterns and make the necessary adjustments in order to minimize the differences between annotators.

[§] The authors wish it to be known that, in their opinion, the first two authors should be regarded as joint First Authors.

Neves and Leser (9), and Neves and Seva (10) have recently provided extensive reviews for automatic annotation tools (11-22) and, as a result of their analysis, they identify a list of criteria that maximize a tool's use for annotation/curation of gold standard corpora. These criteria are: 1) technical – users prefer tools that are publicly available, web-based and open source, with an option for local installation to allow for the secure annotation of documents such as clinical records, 2) data – users prefer that tools handle the standard formats for input/output of documents and annotations and can be easily applied to PubMed, 3) functional – users prefer tools that handle multi-label annotations, document-level annotations, relational annotations, full text annotations, as well as multiple-user and team annotations. In addition, the ideal tool should allow multiple languages, support links to ontologies and terminologies, and provide for quality assessment and inter-annotator agreement calculations. Finally, the authors evaluate a tool's suitability for biomedicine by its ability to support the integration with PubMed or PMC, as this facilitates the retrieval, parsing, and even pre-processing of documents for further processing and annotation.

In this paper, we introduce *TeamTat*, a web-based, open-source, collaborative text annotation tool equipped to manage the production of high-quality annotated corpora, fulfilling all of the major criteria listed above and more. In short, TeamTat features 1) full-text support showing the document in its entirety including figures as they are an integral part of manual biomedical annotation/curation; 2) easy integration with PubMed and PMC through BioC, a simple format for sharing text data and annotations towards improved interoperability created by the text mining research community; 3) an intuitive and user-friendly interface for all users to review and analyse their annotations, independently and collaboratively; and 4) a quality assessment and management mechanism to bring this all together from a project administration perspective. Taken together, TeamTat is an all-in-one system with a set of features that cannot be found in existing tools. For example, our previous tool, ezTag (18), does not support team annotation and project management. Similarly, other tools surveyed in (9,10) are limited in their support of local installation, PubMed/PMC integration, full-text annotation, figure display, collaborative annotation, etc.

In the first release of the software, annotation functionalities centred on tagging of entity or concept mentions, enabling the flexible definition of entity classes/types, making the annotation as easy as possible for human curators and supporting annotation quality analysis at the entity level. We further added the document triage functionality, and the ability to define and annotate relations between entities, which are document level relationships and therefore not confined to a single sentence or even a single paragraph. Finally, we added project management functionality and inter-annotator agreement statistics. *TeamTat*'s original design was based on our prior experience in developing various biomedical corpora. It has since been used in several projects at the National Library of Medicine (NLM) involving annotation of genes and chemicals in PubMed and PubMed Central articles (see Use Case). These experiences have enabled the identification of desirable refinements and extensions.

The aim of this paper is to describe *TeamTat*'s open source platform and the annotation and analysis perspectives that make it an easy-to-use flexible tool for biomedical document annotation and curation. The following sections provide technical details about software development and showcase the functionalities through demonstrations inspired by creating public biomedical corpora. Throughout this paper, the terms: annotation and curation, annotator and curator are used interchangeably. While *TeamTat* can be used both for triage and full curation tasks, in this paper we focus on the description of *TeamTat* features helping text annotators, database curators, and project managers make it possible to create richly annotated gold standard corpora.

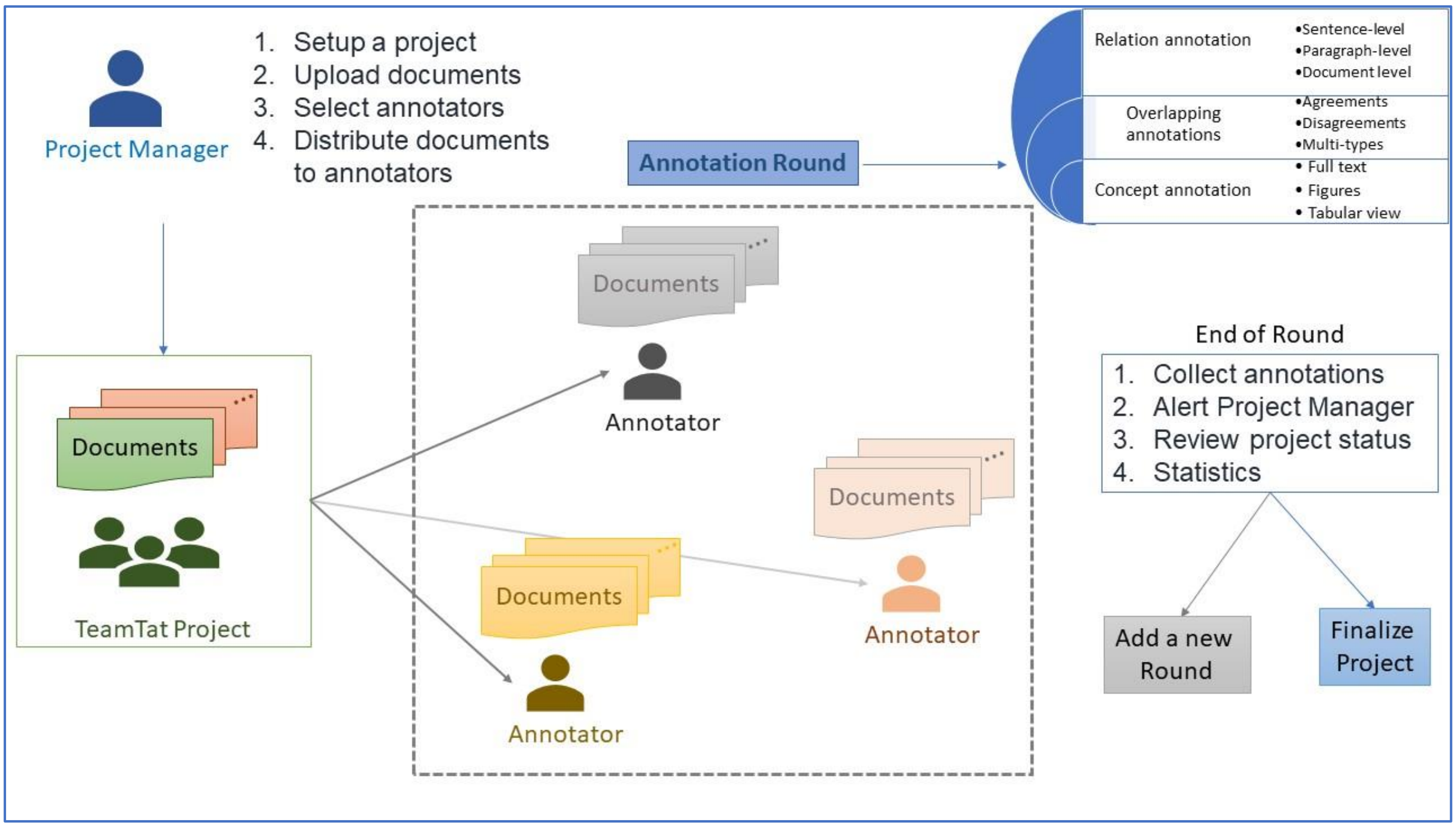

**Figure 1 Overview of a *TeamTat* annotation project.** A project manager selects the documents to be annotated, specifies the types of entities and relations to be considered, enables the participation of the annotator(s), and distributes the documents among annotators. An annotation round consists of team members working independently to annotate. Team members can also review annotations where they have a disagreement among annotation partners. *TeamTat* maintains anonymity to prevent annotator bias. At the end of each round, the project manager calculates inter-annotator agreement statistics, and decides to continue or finalize the corpus. Annotations are trackable for every annotation round, and data can be downloaded at any time.

## SYSTEM DESCRIPTION

Collaborative text annotation is a complex process, and requires domain experts, project managers and a wide range of automatic pre-processing, user interface, and evaluation tools. *TeamTat* offers multiple-user roles, and provides administrative, project management and annotation interfaces.

The annotation interface is designed for ease of integration with PubMed/PMC, ease of use with full text articles and supports figure display for the annotators' convenience. Annotators can work on the same document independently and anonymously in their workspaces, or collaboratively, with live discussions to resolve disagreements. The important features of the annotation interface in *TeamTat* include: 1) annotators may collaborate, 2) annotators can annotate documents of any length, including full text journal articles, 3) the *TeamTat* interface can display all figures of PubMed Central full text articles, 4) documents can be added from the PubMed/PMC BioC APIs (23,24), or uploaded from local repositories, 5) the annotation interface is optimized for user-friendly browsing, 6) documents can be annotated for triage, for entities and for relations.

The project management interface allows: 1) annotation projects to be organized in multiple rounds, 2) project managers to track task development and completion, 3) the assessment of annotation quality via inter-annotator agreement statistics, and 4) the creation of corpus report statistics. Figure 1 presents the overall annotation workflow of a given project. The administrative functions include all of the above. Administrators can set up *TeamTat* locally to accommodate data privacy concerns. Documents can be in BioC, plain text or PDF format, and Unicode support allows for documents in different languages. In the case of a PDF file, text is automatically extracted using Docsplit ruby library (https://rubygems.org/gems/docsplit). Annotation data can be readily downloaded and exported in BioC, which is interoperable to other formats such as PubAnnotation-JSON (25,26).

### Implementation

We developed *TeamTat* using Ruby on Rails and MySQL as a backend database. All the web pages are HTML5/CSS compatible, thus it supports the latest version of popular web browsers (e.g. Chrome, Safari, Firefox, Internet Explorer) and mobile devices. The source code is available at [https://github.com/ncbi-nlp/TeamTat].

## USAGE

*TeamTat* offers support for annotation efficiency, consistency, scale; provides an intuitive interface; and mimics a project development workflow with clear procedures that allow the development of a gold standard corpus. *TeamTat* does not collect any personal information data from its users.

### Project Management Features

The project manager can define a curation project, customizing it as needed based on the project requirements (annotation guidelines). The project manager selects the documents to be annotated, specifies the types of entities and relations to be considered, and enables the participation of one or more annotators in the project. An annotation round reflects the iterative nature of the production life cycle. The initial round may consider unannotated documents or pre-annotated documents using an external automated system. *TeamTat* facilitates independent work of multiple annotators on the same documents and provides an environment to evaluate the quality of annotations via inter-annotator agreement statistics. Multiple annotation rounds are recommended to ensure a high-quality corpus.

Annotation rounds can be individual or collaborative. An individual annotation round allows each annotator to work in their individual workspace, and review/edit/revise annotations. The identity of the annotation partners on the same document is kept hidden to guard against bias, while the agreements and disagreements between partners are displayed via visual cues. A collaborative annotation round is usually the last review before finalizing the project. During this round, the identities of annotation partners are revealed, so that they may discuss any remaining discrepancies, and agree on how to resolve them.

A key feature to facilitate the project managers' work in *TeamTat*, is the ability to assess corpus quality via inter-annotator agreement statistics. At any time during the annotation project, the project manager is able to track via their interface the agreements and disagreements between annotators. Once an annotation round is complete, the project manager can calculate exact agreement between annotators (requiring complete agreement on annotation type, span, and concept normalization) as well as different levels of soft agreement (accounting for partial overlap, mismatched types, or differences in concept ID). Once corpus quality is considered acceptable (administrative decision), the final corpus is produced. All data is available for download at any time.

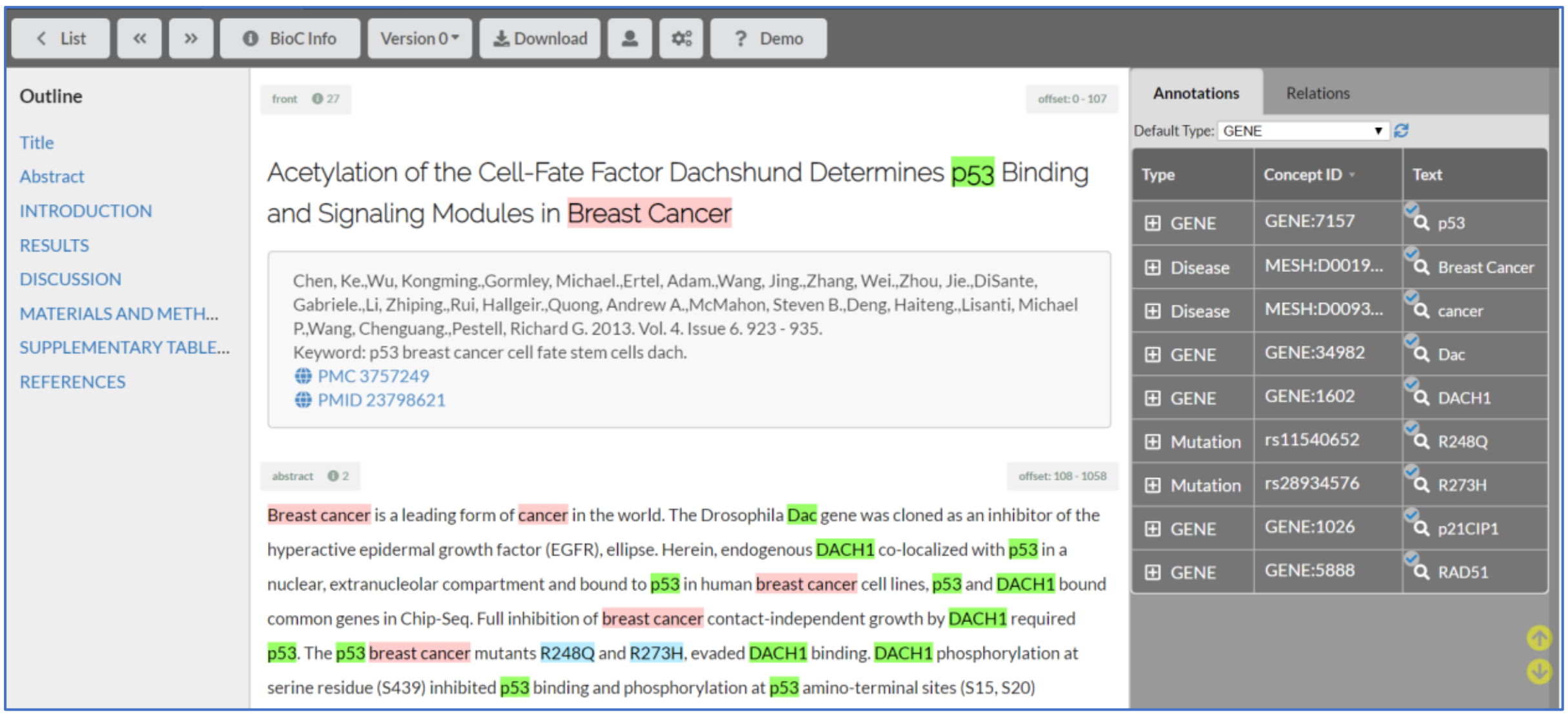

**Figure 2. Screenshot of the *TeamTat* annotation editor.** The middle of the screen shows the article content. Title, and metadata are listed first, as well as automatic links to the PMID and PMCID records respectively. The right-hand side shows curated entities and relations.

### Annotation Features

Motivated by the experience of annotation projects during BioCreative V (27,28) and VI (1,29), and feedback from PubTator Central (30) and ezTag (18) users, *TeamTat* was designed to include useful features from ezTag, PubTator Central, and Marky (15,16,31). As such, *TeamTat* offers an improved smart interface, full text support, and relation annotation.

*TeamTat's* annotator user interface is very intuitive. Once an annotator is assigned to a project, they will find the project listed in their workspace. A project typically contains several documents, which can be annotated in any random order. In addition to the annotation editor, annotators also have a document list view, which allows them to see a summary of the assigned documents and their annotation status such as: the number of annotations per document, their assigned triage label, completion status, and the time of the last update. Documents can be sorted, and a search function is available to retrieve any document matching a keyword from the current collection.

The annotation editor (Figure 2) is a smart interface that aims to minimize the number of actions required to add/delete or revise annotations. Annotations can be character level, word level or phrases, and are automatically saved as they are highlighted. Overlapping annotations are allowed, and they can be different entity types. All occurrences of the highlighted text in the given document can be annotated/edited simultaneously. The entity type has a drop-down box listing all the annotation types (or tags/categories) valid for the current project. These are generated automatically from the annotation schemas defined by the project manager. Annotation schemas define the acceptable range of annotations and thus allow the user interface to be customized. Since *TeamTat* is a general annotation tool, only manual typing is allowed for entering concept IDs. The last step of manual annotation is to toggle on the 'Done' button, which indicates that the annotation is complete, and the annotator moves on to the next document.

One of the desiderata in biomedical annotation tools is the capability to annotate relations between entities. Relations can be between entities of the same type (such as genetic interactions, protein-protein interactions) or between different types of entities (such as gene-disease relations, e.g. Figure 3). *TeamTat* ensures that entity types selected by the user are consistent with the annotation schema defined by the project manager. Moreover, *TeamTat* relations are not restricted to binary relations. They can have up to eight components, and they are at the document level, meaning that individual nodes are not restricted to appear in the same sentence, or even in the same paragraph.

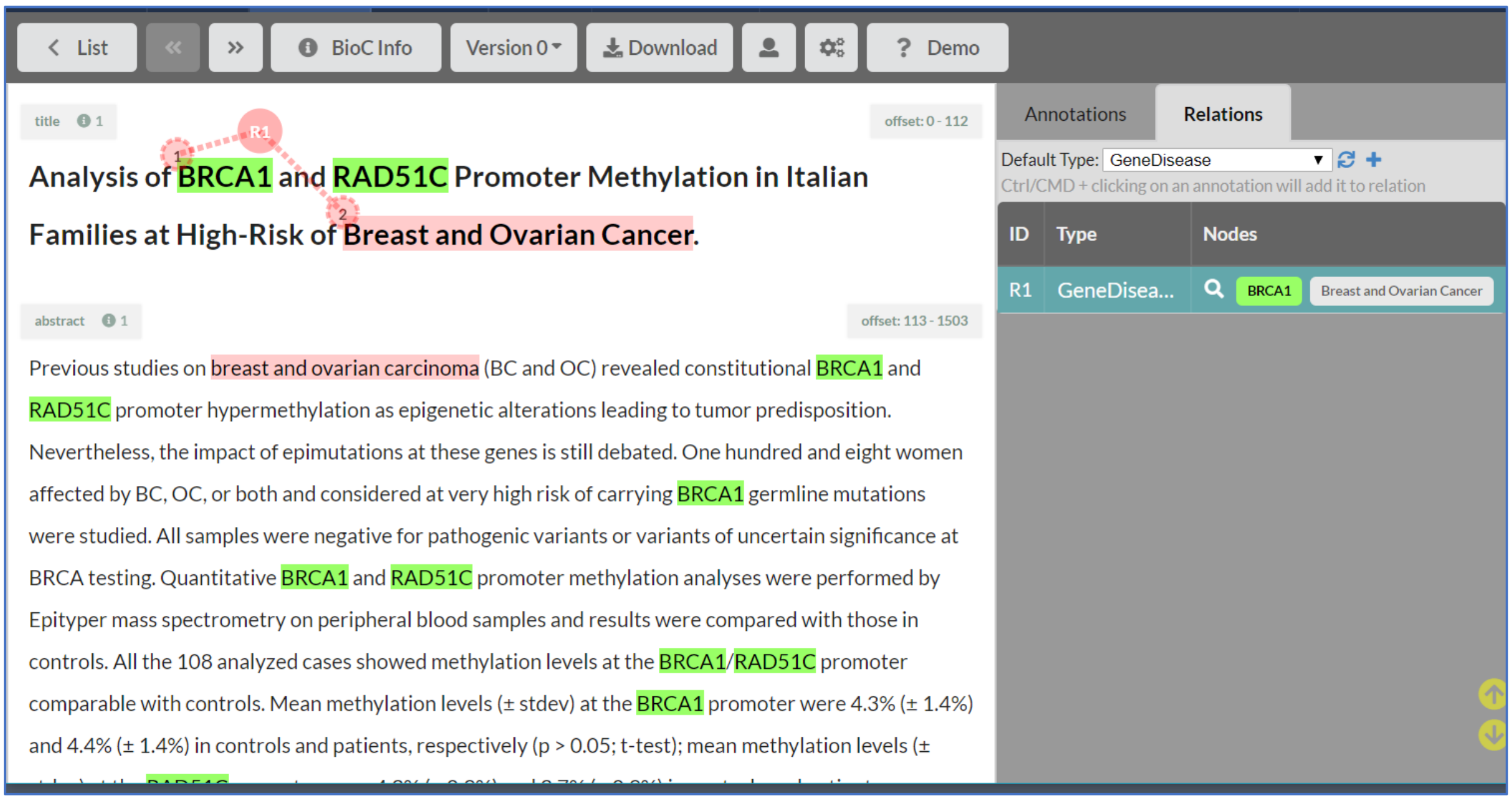

**Figure 3. Relation annotation in *TeamTat*.**

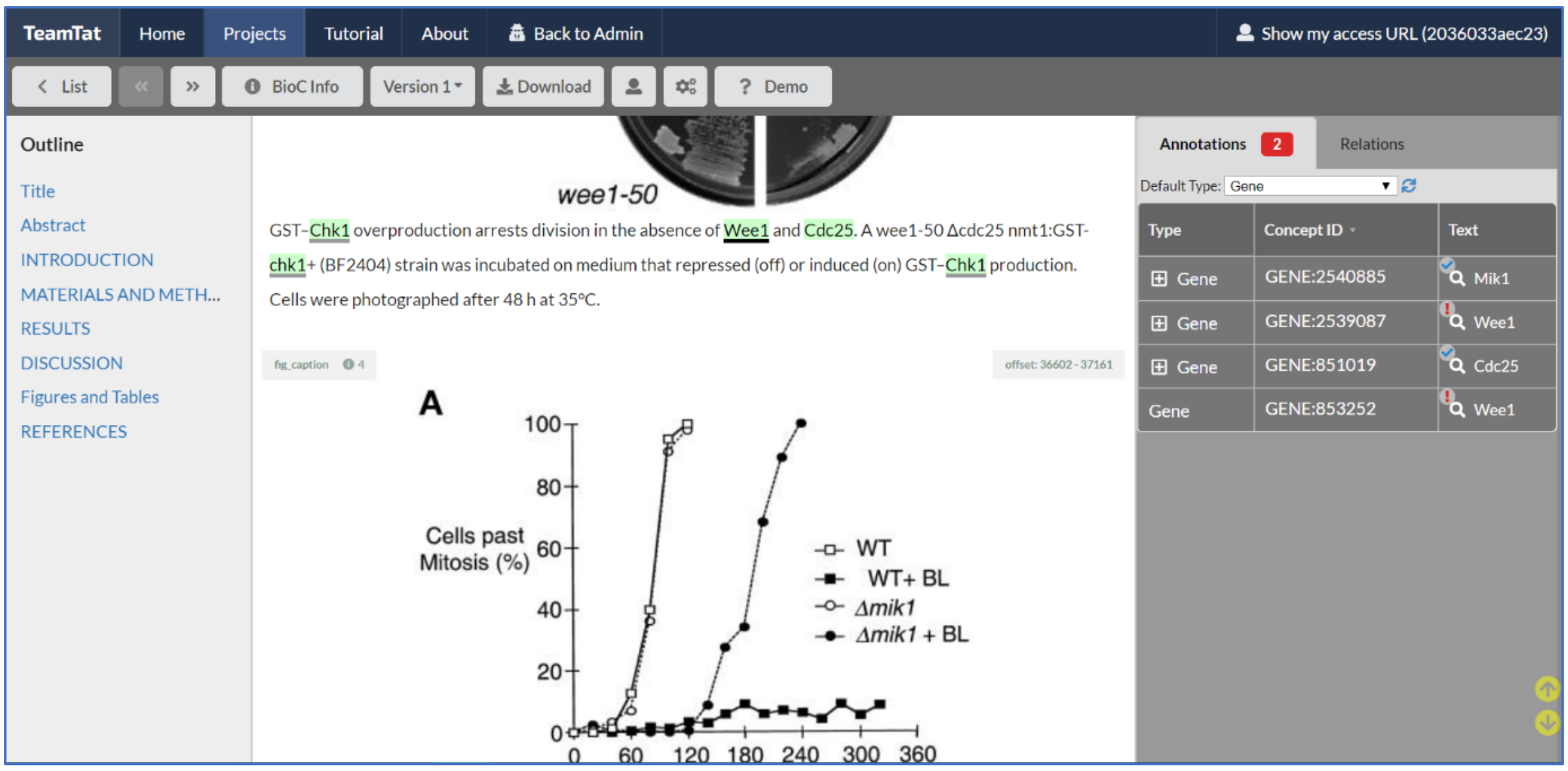

**Figure 4.** ***TeamTat*** **incorporates visual cues to alert for agreements and disagreements among annotators.** In this figure, we see that the gene "Chk1" has a grey underline, "Wee1" has a black underline, and "Cdc25" has no underline. These cues provide the project manager with the information that annotators agree on the annotation of "Chk1", they disagree on the Concept ID for gene "Wee1" and only one annotator has annotated "Cdc25". Furthermore, this figure shows how *TeamTat* incorporates the display of images in full text articles to facilitate the work of annotators.

## Collaborative Features

One of the most convenient features of *TeamTat* is the intuitive full functionality environment that allows for both independent work, as well as collaborative work between annotators. One major consideration in human data annotation is to control for bias, and *TeamTat* allows the project manager to pair the annotators anonymously and distribute the data anonymously to their individual workspaces, during the independent annotation rounds. The first annotation round typically consists of annotators working independently on their assigned documents (which may be raw or pre-annotated). At this stage, while the project manager can see all edits in real time, annotators cannot see the changes on the same document performed by other annotators. If the project manager decides to follow up with an additional independent review round (e.g. Figure 4), each annotator can see in their workspace the agreements and disagreements, while the identity of the working partners is still kept hidden to allow for an unbiased review and revision stage. The project manager can repeat this iterative process until they are satisfied with the results. *TeamTat* also provides a convenient collaborative discussion annotation round, in which identities of annotation partners are revealed, and they are encouraged to finalize their discrepancies in a collaborative way. We recommend this mode of operation during the preparation phase of annotation guidelines, as well as the final annotation round for full curation projects.

## USE CASE

*TeamTat* has been used for the development of two recent corpora at the National Library of Medicine, the NLM-Chem corpus, a collection of 150 full text articles annotated for chemicals, and the NLM-Gene corpus, a collection of 550 PubMed articles annotated for genes from 11 model organisms. The NLM-Chem corpus was doubly annotated by ten experienced NLM MeSH indexers, and the NLM-Gene was doubly annotated by six NLM MeSH indexers. Both sets of articles were selected to be highly ambiguous, rich in biomedical entities and from a large variety of journals in the PMC Open Access subset. *TeamTat* provided the right environment to manage these heavy annotation loads, providing the project manager with the right functionality to balance the annotation load amongst annotators, and allow room for both independent work, and weekly discussion meetings.

Most of the tool functionality was improved as a result of the frequent interactions with the annotators and implementation of their recommendations. The annotators appreciated the ease of annotation, the ability to conveniently interact with PubMed and PMC, and the ability to see the figures within the display, since figures frequently contain crucial information or experimental evidence. For the NLM-Chem corpus the ability to navigate the full text via the left-side panel, and the tool's ability to remember the last paragraph where they left off were highly appreciated. For both projects, the right-side tabular list of annotated entities provided the ability to work on one entity at a time, as opposed to the middle panel which provided the sequential text access to the document.

Since both NLM-Chem, and NLM-Gene aimed to provide complete and thorough annotations for all mentioned chemicals and genes, they were projects with heavy annotator involvement. Therefore, both projects went through 4 or 5 annotation rounds, until the all disagreements and discrepancies were resolved. During these discussion and revision stages, *TeamTat* provided a rich environment to review annotations, both in the independent rounds, as well as in the collaborative final rounds. The indexers appreciated the visual cues to alert them to: the disagreements on annotation span, versus normalization, and missed annotations noted by only one of the working partners. Likewise, they also appreciated the automatic links to the corresponding records in NCBI GENE and MeSH for each linked entity. Both corpora will be released to the research community to foster better recognition of genes and chemicals in scientific publications.

## CONCLUSIONS

We have described *TeamTat*, a web-based system for collaborative text annotation, and management of multi-annotator projects. *TeamTat* supports annotation of both entities and relations and is integrated with PubMed and PMC. Document input format can be plain text, PDF or BioC XML, and output documents are BioC XML with inline annotations. In addition to the web-based system, *TeamTat* can also be set up locally.

*TeamTat* is a web-based, open-source, collaborative document annotation tool equipped to manage the production of high-quality annotated corpora. *TeamTat* realizes the desired criteria for a complex annotation process, which involve: 1) interactive, intuitive user interface supporting documents of any length, including full text articles and figure display, to improve annotation efficiency, 2) support for pre-annotation to help achieve time and cost savings, 3) support for both entity and relation annotation, with the ability to adapt to different annotation guidelines, 4) multi-role support, including annotator, project manager and administrator, and corresponding user interfaces, 5) support for corpus quality assessment, and the ability to organize annotations in a multi-round process until desired corpus quality is achieved.

*TeamTat* has been active since March 2019 for team annotation projects at NLM, and word-of-mouth has generated a large active base of more than a hundred document annotators. We expect *TeamTat* to become an important tool for development of gold standard corpora in biomedicine and beyond, because it provides the right blend between project manager, domain expert annotator and a collaborative environment.

## AVAILABILITY

*TeamTat* is an open source web-based annotation tool, available at https://www.teamtat.org and in the GitHub repository [https://github.com/ncbi-nlp/TeamTat].

## ACKNOWLEDGEMENT

This research is supported by the NIH Intramural Research Program, National Library of Medicine. We thank TeamTat users for their helpful feedback and suggestion for its enhancements.

## FUNDING

This work was supported by the National Institutes of Health, Ministry of Science and ICT (MSIT) [NRF-2014M3C9A3064706 to D.K.]; Ministry of Education [NRF-2018R1D1A1B07044775 to D.K.]; Funding for open access charge: National Institutes of Health.